\documentclass[preprint,12pt,authoryear]{elsarticle}

\usepackage{amssymb}
\usepackage[flushleft]{threeparttable}
\usepackage{rotating}
\newcommand{\kms}{\hbox{km\,s$^{-1}$}}
\newcommand{\mum}{$\mu$m}
\newcommand{\teff}{$T_{\rm{eff}}$}
\newcommand{\logg}{$\log g$}
\newcommand{\lL}{\ifmmode \log \frac{L}{L_{\odot}} \else $\log \frac{L}{L_{\odot}}$\fi}
\newcommand{\mdot}{$\dot{M}$}
\newcommand{\myr}{M$_{\odot}$ yr$^{-1}$}
\newcommand{\vsini}{$V$~sin$i$}
\newcommand{\vinf}{$v_{\infty}$}

\newcommand{\msun}{M$_{\odot}$}

\DeclareRobustCommand{\ion}[2]{\textup{#1\,\textsc{\lowercase{#2}}}}
\newcommand*\element[1][]{%
  \def\aa@element@tr{#1}%
  \aa@element
}

\journal{New Astronomy}
\begin{document}

\begin{frontmatter}



\title{Properties of massive stars in four clusters of the VVV survey \tnoteref{label1}}
\tnotetext[label1]{Based on observations with ISAAC/VLT/ESO (programme 087.D-0341A), New Technology Telescope at ESO/La Silla Observatory (programme 087.D-0490A), VVV ESO Large Survey (programme 179.B-2002) and with SOAR (programme CN2012A-616).}

\author[1,2]{A. Herv\'e}
\ead{anthony.herve@asu.cas.cz}
\author[2]{F. Martins}
\author[3]{A.-N. Ch\^en\'e}
\author[4]{J-C. Bouret}
\author[5,6,7]{J. Borissova}

\address[1]{Astronomical Institute ASCR, Fri\v{c}ova 298, 251 65 Ond\v{r}ejov, Czech Republic}
\address[2]{LUPM, Universit\'e de Montpellier, CNRS, Place Eug\`ene Bataillon, F-34095 Montpellier, France}
\address[3]{Gemini Observatory, Northern Operations Center, 670 N. A'ohoku Place, Hilo, Hawaii, 96720, USA}
\address[4]{Aix Marseille Universit\'e, CNRS, LAM (Laboratoire d'Astrophysique de Marseille) UMR 7326, 13388, Marseille, France}
\address[5]{Instituto de F\'isica y Astronom\'ia, U. de Valpara\'iso, Av. Gran Breta$\tilde{n}$a 1111, Playa Ancha, 5030 Casilla, Chile}
\address[6]{The Milky Way Millennium Nucleus, Av. Vicu$\tilde{n}$a Mackenna 4860, 782-0436, Macul, Santiago, Chile}
\address[7]{Millenium Institute of Astrophysics, MAS, Av. Gran Bretana 1111, Playa Ancha, Casilla 5030, Valparaiso, Chile}

%-----------------------------------------------------------------------

\begin{abstract}
The evolution of massive stars is only partly understood. Observational constraints can be obtained from the study of massive stars located in young massive clusters. The ESO Public Survey ``VISTA Variables in the V\'ia L\'acte\'a (VVV)'' discovered several new clusters hosting massive stars. We present an analysis of massive stars in four of these new clusters. Our aim is to provide constraints on stellar evolution and to better understand the relation between different types of massive stars. We use the radiative transfer code CMFGEN to analyse K-band spectra of twelve stars with spectral types ranging from O and B to WN and WC. We derive the stellar parameters of all targets as well as surface abundances for a subset of them. In the Hertzsprung-Russell diagram, the Wolf-Rayet stars are more luminous or hotter than the O stars. From the log(C/N) -- log(C/He) diagram, we show quantitatively that WN stars are more chemically evolved than O stars, WC stars being more evolved than WN stars. Mass loss rates among Wolf-Rayet stars are a factor of 10 larger than for O stars, in agreement with previous findings. 

\end{abstract}
%-----------------------------------------------------------------------


\end{frontmatter}
%-----------------------------------------------------------------------

\section{Introduction}

Through their powerful stellar winds, their strong UV radiation field and their violent end of life as supernova explosion, massive stars play a key role in the ecology and in the structure of the galaxies. They are major agents of galactic chemical enrichment and they give birth to new generations of stars by triggering star formation through their feedback effects \citep{deharveng05}.

Nevertheless, even-though their global evolution scheme as a function of initial mass is known \citep[e.g.][]{crowther07}, the details are not well understood. At solar metallicity O-type stars with an initial mass above 40-60 \msun\ evolve into H-rich WN stars, and Luminous Blue Variables (LBV), before turning into H-poor WN stars and finally WC stars \citep{crowther07}. For O-type stars with lower initial masses, the H-rich WN and LBV phases are replaced by a blue and a red supergiant phase, respectively. But the detailed evolutionary path of a star born with a given mass is not accurately determined. Indeed, numerous physical effects (rotation, mass loss rate, magnetic field, binarity) strongly impact the internal structure of massive stars and thus their evolution. Evolutionary calculations allow to study these effects and predict the life and death of massive stars. But such calculations rely on various assumptions \cite[e.g.][]{martins13}.

In this context, the study of young massive clusters containing WR stars is a key to understand the detailed evolution of massive stars.
Indeed, if one assumes coeval star formation in massive clusters, the comparison of evolved and main sequence stars allow us to constrain the nature of the most massive and evolved objects (their mass being higher than main sequence objects still present in the cluster, since more massive stars live shorter). For instance \cite{martins07} showed that in the central cluster of the Galaxy, direct connections between O-type and WR stars could be established. More important, the identification of {main sequence} O-type stars is a crucial piece of information: it shows the position of the turn-off of the entire population, providing constraints on the age \cite[e.g.][]{paumard06} and consequently an upper limit on the main sequence life time of the WR's progenitors. The presence of red supergiants and WR star(s) in the same cluster constraint the link between hot and cool massive stars, and the evolutionary sequence between red supergiants and WR stars -- Westerlund 1 {\citep{clark05}} and central cluster \citep{paumard14}.

Because they are rare, far away and thus located behind lots of dust, young massive clusters cannot be observed at UV and optical bands where extinction is too strong. The infrared (IR) VISTA Variables in the V\'{\i}a L\'actea (VVV) survey \citep{Mi10,Sa12} provides hundreds of embedded clusters near-infrared photometry of tens of embedded clusters \citep{Bo11,Bo14,solin14,barba15}. \cite{Bo11} discovered 96 new open clusters and stellar groups in the Galactic disk area. In this sample, six heavily-obscured clusters contain at least one WR star together with several main sequence OB stars \citep{chene13}. Although the census of the clusters' population is not complete, these clusters likely sample various stellar ages and thus different phases of massive star evolution. Studying their stellar content is thus a first step towards a global characterization of stellar evolution at high masses.

In this paper, we perform a quantitative spectroscopic analysis of massive stars in four of these six clusters. We determine their fundamental parameters and, when possible, their surface abundances. 
In Sect.\ \ref{s_obs}, we summarize the observations and data reduction. We describe in Sect.\ \ref{s_mod} our modelling tools and analysis strategy. The results are described in Sect.\ \ref{s_res} and discussed in Sect.\ \ref{s_disc}. Conclusions are given in Sect.\ \ref{s_conc}.

%%###############################################
%%###############################################
\section{Sample, observations and data reduction}
\label{s_obs}

The observations and data reduction were already presented in \cite{chene13}. They are briefly summarized below.

Spectra were collected on different instruments: ISAAC on the VLT at ESO/Paranal Observatory; SofI on the NTT at ESO/La Silla Observatory; OSIRIS on the SOAR telescope. Total exposure times were typically 200-400 s for the brightest stars and 1200 s for the faintest. Details are given in \cite{chene13}.

All reduction steps were executed using both custom-written Interactive Data Language (IDL) scripts and standard \emph{iraf4} procedures. Subsequent nodding observations were subtracted from one another to remove the bias level and sky emission lines.
Flat fielding, spectrum extraction and wavelength calibration of all spectra were done in the usual way using \emph{iraf}. Calibration lamp spectra (helium-argon and neon for OSIRIS, xenon-neon for SofI and ISAAC) taken at the beginning of each night were used for wavelength calibration. Correction for atmospheric absorption was done using the \emph{iraf} task \emph{telluric}.  Finally, all spectra were rectified using a low-order
polynomial fit to a wavelength interval that was assumed to be pure continuum, i.e. where no absorption or emission lines were
observed nor expected.  Due to some problems with the data, the ISAAC pipeline could not be used. A manual reduction was done. More details on the problems and solutions adopted are detailed in \cite{chene13}.

\begin{table*}
\begin{threeparttable}
\caption{Position, photometry and spectral parameters of the spectroscopic targets.}
\label{TabSp}
\centering
\begin{tabular}{rrccrrrcccc}
\hline\hline
\noalign{\smallskip}
&\multicolumn{1}{c}{ID}&      R.A.          &      Dec       &\multicolumn{1}{c}{$J$}&\multicolumn{1}{c}{$H$}&\multicolumn{1}{c}{$K_{\rm s}$} & {Sp. Typ.} & MK\\
&            &   (J2000)     &   (J2000)    &                &              &                &                  &  \\
\noalign{\smallskip}
\hline
\noalign{\smallskip}
\multicolumn{4}{l}{\it VVV\,CL009 (d= 5 $\pm$ 1 kpc)} &&&&&&&\\
& 5  & 11:56:03.07 & --63:18:54.63 & 11.87 & 11.31 & 11.04 & O8--9V         &  -3.08 \\
& 6  & 11:56:03.78 & --63:18:54.44 & 10.55 &  9.93 &  9.65 & OIf/WN7        &  -4.51  \\
& 8  & 11:56:04.38 & --63:18:54.44 & 12.14 & 11.59 & 11.34 & O8--9V         &  -2.76  \\
%\multicolumn{11}{l}{}\\
%\multicolumn{4}{l}{\it VVV\,CL036 (d= 2 $\pm$ 1 kpc):}&&&&&&&\\
%& 4  & 14:09:02.75 & --61:15:58.94 & 14.25 & 10.95 &  9.32 & 2--3I          &  -5.27  \\
%& 9  & 14:09:04.30 & --61:15:53.78 & 13.37 & 10.66 &  9.16 & WN6            &  -5.00  \\
\multicolumn{11}{l}{}\\
\multicolumn{4}{l}{\it VVV\,CL073 (d= 4 $\pm$ 1 kpc):} &&&&&&\\
& 2  & 16:30:23.73 & --48:13:04.90 & 15.40 & 12.51 & 11.08 & O4--6If+/WN9   &  -4.65  \\
& 4  & 16:30:23.98 & --48:13:05.48 & 14.71 & 11.53 &  9.92 & WN7            &  -6.09  \\
\multicolumn{11}{l}{}\\
\multicolumn{4}{l}{\it VVV\,CL074 (d= 6 $\pm$ 1 kpc):}& &&&&&&\\
& 1  & 16:32:05.24 & --47:49:29.13 & 16.40 & 13.53 & 12.09 & O8.5I          &  -4.51  \\
%& 2  & 16:32:05.27 & --47:49:14.25 & 16.92 & 13.23 & 10.31 & WC8            &  -7.67 / -5.00*  \\
& 3  & 16:32:05.46 & --47:49:28.10 & 14.72 & 11.82 & 10.17 & WN8            &  -6.58  \\
& 5  & 16:32:05.49 & --47:49:29.80 & 15.64 & 12.90 & 11.54 & O8.5I          &  -4.94  \\
& 7  & 16:32:05.67 & --47:49:30.00 & 15.98 & 13.46 & 12.19 & O8.5I          &  -4.10  \\
& 9  & 16:32:05.93 & --47:49:30.92 & 15.22 & 12.53 & 11.31 & O4--6If+       &  -5.05  \\
\multicolumn{11}{l}{}\\
\multicolumn{4}{l}{\it VVV\,CL099 (d= 4 $\pm$ 1 kpc):} &&&&&&\\
& 5  & 17:14:25.42 & --38:09:50.40 & 13.43 & 11.70 & 10.64 & WN6            &  -4.17  \\
& 7  & 17:14:25.66 & --38:09:53.72 & 11.67 & 10.06 &  9.26 & WC8            &  -5.32  \\
\noalign{\smallskip}
\hline
\end{tabular}
\begin{tablenotes}
  \footnotesize
  \item Columns include star's name, position (R.A. and Dec), $J$, $H$ and $K_{\rm s}$ photometry, $E(J-K)$ and $E(H-K)$ colour excesses, extinction ($A_{K}$) from \cite{chene13}. Spectral types have been refined compared to \cite{chene13}.
\end{tablenotes}
\end{threeparttable}
\end{table*}

In Table \ref{TabSp} we present the observational properties of the selected stars. The classification of the different stars, based on the catalogs of \cite{hanson96}, \cite{hanson05} and \cite{figer97}, have been presented in \cite{chene13}. They were refined for the O stars in cluster VVV\,CL074. We also provide the absolute K-band magnitudes. They have been computed using $(J-K)_{0}=-0.21$ and $(H-K)_{0}=-0.10$ \citep{martinsplez06} and $A_{K}=0.6 \times\ E(j-K)$. We checked that the resulting absolute magnitudes corresponded to the calibrations of \citet{martinsplez06} for O stars, and \citet{rosslowe} for Wolf-Rayet (WR) stars. In one case -- star VVV\,CL074-2 -- we found a large discrepancy between the calculated and tabulated values: this WC8 star has MK=-7.67 while the average value for such objects is about -5.0 \citep{rosslowe}. This may be explained by the presence of dust emission contributing to the K-band flux. We noted that the line intensities for that object are quite small, favouring this interpretation. We thus decided to exclude VVV\,CL074-2 from our study.

Compared to the sample of \citet{chene13} we have also excluded VVV\,CL099-04. The spectrum of this star clearly shows the presence of O-type absorption lines superimposed on top of broad WN-type emission lines, hinting at a binary nature. As we do not have multi-epoch spectra we are not able to disentangle the contribution on each star on the combined spectrum. Consequently we can not perform an analysis of stellar wind parameters for this system. 
We did not consider VVV\,CL036-9, a WN6 star, since its line intensities are about 50\% lower than that of normal WN6 stars \citep[e.g.][]{figer97}. Test models revealed that it was not possible to reproduce both the line intensities and the absolute magnitude of the star. \citet{borissova14} showed that VVV\,CL036-9 was surrounded by a dust shell that may explain the morphology of spectral lines. 

Finally, we also excluded main sequence B stars from our sample. These stars present only one hydrogen line in the K-band (Br$\gamma$) from which it is not possible to derive a temperature, surface gravity and other stellar parameters.
Finally we ruled out stars for which the data quality (low signal-to-noise ratio) was not sufficient to perform a spectroscopic analysis (as VVV\,CL011-2).
As a consequence, their are no stars available for a study in two of the massive clusters presented in \cite{chene13} VVV\,CL036, and VVV\,CL011.

%%###############################################
%%###############################################
\section{Spectroscopic analysis}
\label{s_mod}

We used the radiative transfer code CMFGEN \citep{HillierMiller98} to determine the stellar and wind parameters of the massive stars presented in the previous section. For a spherically-symmetric and steady-state outflow, this modeling tool calculates continuum and line formation in the non-LTE regime by solving the coupled statistical equilibrium and radiative transfer equations. The main input parameters are: the effective temperature $T_{\rm eff}$, the luminosity $L_{*}$ (or radius $R_{*}$), the mass (or surface gravity), the mass-loss rate $\dot{M} $, the wind terminal velocity $ v_{\infty} $ as well as the exponent of the wind-velocity slope (the so-called $\beta$ parameter, set to 1.0 in our calculations), and the abundances of individual elements. We have used the following elements in our calculations: H, He, C, N, O, Ne, Si, S, Fe, Ni. A total of 6215 energy levels (treated using a super-level approach) and about 150 000 transitions were used in our models. The reference solar abundances were adopted from \citet{ga10}. Finally, wind clumping \citep{ELM,BLH,Puls08} was implemented as follows: 

\begin{equation}
 f(r)=f_{\infty}+(1-f_{\infty})*exp[-\frac{v(r)}{v_{c} }]
\end{equation}

\noindent where $f$ is the volume filling factor, $f_{\infty}$ is the value of $f$ at the top of the atmosphere, $v(r)$ is the velocity of the wind and $v_{c}$ is the velocity where the wind become inhomogeneous. We adopted $f_{\infty} = 0.1$ and $v_{c} = 100$ \kms\ in our calculations. 

Once the atmosphere model was converged, a formal solution of the radiative transfer equation was performed to produce the emergent spectrum. In that step, detailed line profiles including Stark broadening were taken into account. Synthetic spectra were subsequently compared to the observations to determine the main parameters. Using spectral lines listed in Table~\ref{obslines}, we proceeded as follows:

\begin{itemize}

\item {\it{Effective temperature}}: $T_{eff}$ was determined from the ratio of the line strength of two successive ionization levels of a given ion. In the the infrared band, we used \ion{He}{i} and \ion{He}{ii} lines (see Tab.\,\ref{obslines}). We excluded \ion{He}{i} 2.058 $\mu$m from the analysis because of its sensitivity to many effects \citep{najarro06}. In addition to $T_{eff}$ which is defined as the temperature of the star where the Rosseland optical depth equals 2/3, we determined $T_*$ as the temperature where the Rosseland optical depth is equal to 20. At these optical depth, the structure of the star corresponds to a quasi hydrostatic layer of the atmosphere, which is more similar to the outer radius of evolutionary models.
In the case of O-type stars, the value of $T_{eff}$ and $T_{*}$ are very close. But in for WR stars, which have dense winds and a large atmosphere, the difference between the two values can reach few thousand of degrees. The typical uncertainty of our measurements is 2000 to 3000 K.\\

\item {\it{Luminosity}}: Once the effective temperature was constrained, we calculated the OB stars K-band bolometric corrections ($BC_{K}$) according to \citet{martinsplez06}. Using the cluster's average distance \citep{chene13} and individual extinction (Table \ref{TabSp}) we then determined the luminosity:

\begin{equation}
log\frac{L}{L_\odot}= -0.4 (M_k +BC_k -M^{\odot}_{bol})
\end{equation}
\noindent where the $M_{K}$ is the absolute K-band magnitude and $M^{\odot}_{bol}$ the Sun's bolometric magnitude. For Wolf-Rayet stars, we set the luminosity from  the absolute K-band magnitude (adjusting \lL\ so that the corresponding $M_{K}$ fits the values from Table \ref{TabSp}). Uncertainties are listed in Table~\ref{tab_param}.

\item {\it{Mass loss rate}}: $\dot{M}$  was determined from the strength of the emission of Br$_\gamma$ and the helium lines. A change of $\dot{M}$ leads to a general increase of the emission in all lines, while an increase of He/H strengthens the He lines and weakens Br$_\gamma$. The uncertainty on $\dot{M}$, all other parameters being held fixed, is about 0.2 dex.\\

\item {\it{Wind terminal velocity }}: $v_\infty$ was determined from the width of Br$_\gamma$ when it is in strong emission and from the extend of the absorption dip of the P-Cygni profile of \ion{He}{i} 2.058 $\mu$m. For the latter indicator, this is a direct measure. The width of Br$_\gamma$ does not give direct access to \vinf, but the higher the terminal velocity, the broader Br$_\gamma$. Hence, models with high \vinf\ produce wider emission lines formed in the wind. When none of these indicators could be used (Br$_{\gamma}$ in absorption; \ion{He}{i} 2.058 $\mu$m contaminated by nebular emission), we simply adopted $v_\infty = 2.6 \times v_{esc}$ \citep{lamers95}, $v_{esc}$ being the escape velocity. \\

\item {\it{Surface abundances}}: the ratio He/H was obtained from the ratio of the strength of Br$\gamma$ and helium lines in emission line stars. Otherwise, it was set to 0.1. Carbon and nitrogen lines (listed in Table\,\ref{obslines}) were sometimes clearly detected. In that case, they were used to constrain the surface abundances, with a typical uncertainty of 70\%.\\

\end{itemize}

We adopted representative values of the surface gravity since the determination of \logg\ was not feasible with our spectra.
In our analysis we assumed $vsini = 100$ km.s$^{-1}$. The resolution of the spectra and their signal to noise ratio do not allow us to use a Fourier transform method to determine more precisely the rotational velocity of our stars. Our synthetic spectra are also convolved with the instrumental resolution of the observed spectra. 

\begin{figure}
\begin{center}
\includegraphics[width=0.9\textwidth]{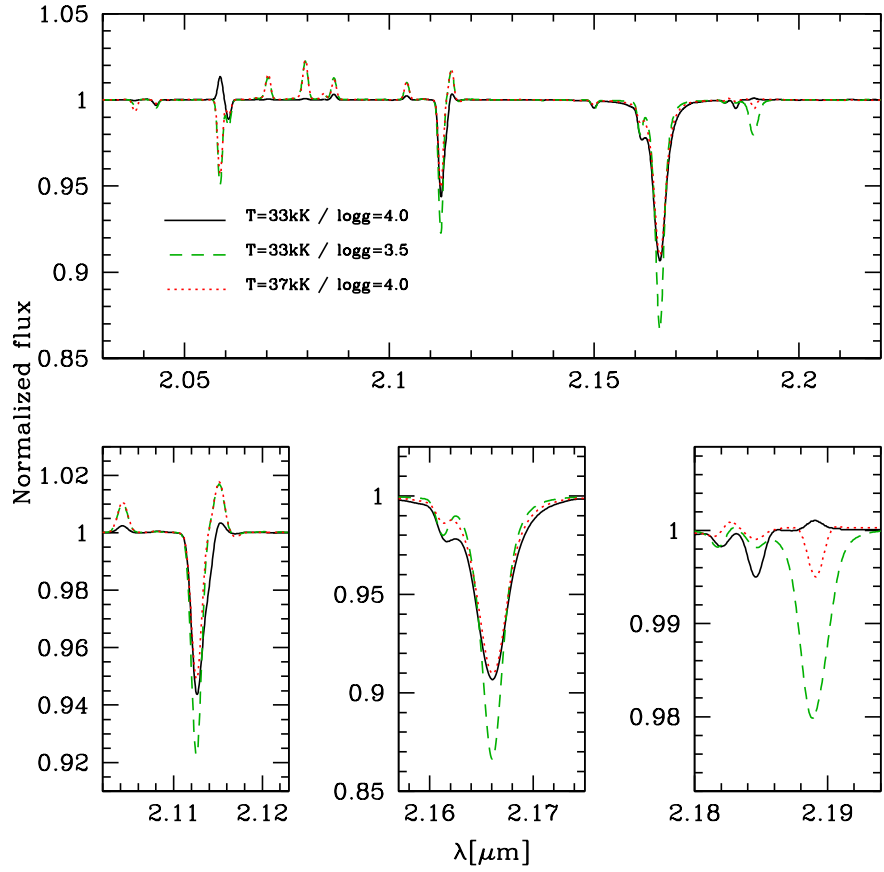}
\caption{Effect of \teff\ and \logg\ on K-band synthetic spectra at a resolution of 2500. The spectra are further convolved with a rotational profile corresponding to \vsini\ = 100 \kms. The stellar parameters, given in the upper panel, correspond to the range of values of the sample's O stars. }
\label{fig_res} 
\end{center}
\end{figure}

We stress that even for the relatively low spectral resolution of our data, the determination of stellar parameters is possible. For Wolf-Rayet stars, the line width does not require higher resolution. For O stars, Fig.\ \ref{fig_res} shows that the effects of \teff\ and \logg\ variations are clearly detected at R=2500. In particular, the \ion{He}{ii}~2.189\mum\ line varies significantly over the range of \teff\ sampled by the OB stars of the present study, ensuring a relatively safe \teff\ determination. 

\begin{table}
\caption[]{$K$-band lines used for the spectroscopic analysis.}
 \label{obslines}
 \begin{center}
 \begin{tabular}{|cl|cl|cl|} 
 $\lambda$ ($\mu$m)   &    Ion  & $\lambda$  ($\mu$m) &    Ion &  $\lambda$ ($\mu$m)   &    Ion       \\  
  \hline
 2.0706      &   \ion{C}{iv} &  2.1155      &   \ion{N}{iii} &  2.1891      &   \ion{He}{ii} \\
 2.0802      &   \ion{C}{iv} &  2.1614      &   \ion{He}{i}  &  2.2471      &   \ion{N}{iii} \\
 2.0842      &   \ion{C}{iv} &  2.1647      &   \ion{He}{i}  &  2.2471      &   \ion{N}{iii} \\
 2.1126      &   \ion{He}{i} &  2.1652      &   \ion{He}{ii} &  2.3142      &   \ion{He}{ii} \\
 2.1138      &   \ion{He}{i} &  2.1661      &   \ion{H}{i}   &  2.3470      &   \ion{He}{ii} \\
 2.1146      &   \ion{C}{iii} \\     
\end{tabular}
\end{center}
\end{table}

%%###############################################
%%###############################################
\section{Results}
\label{s_res}

Our results are presented in Table~\ref{tab_param}. The best fits of the near-infrared spectra are shown in the Fig.\,\ref{fit_vc09},\,\ref{fit_vc73},\,\ref{fit_vc74} and \ref{fit_vc99}. In the following we present a short description of the results for each cluster.

%%###############################################
\subsection{VVV\,CL009 cluster}

Three stars have been analyzed in this cluster: VVV\,CL009-5 and VVV\,CL009-8 are classified as O8-9V and VVV\,CL009-6 as OIf/WN7.

\begin{figure}
\begin{center}
\includegraphics[width=6cm]{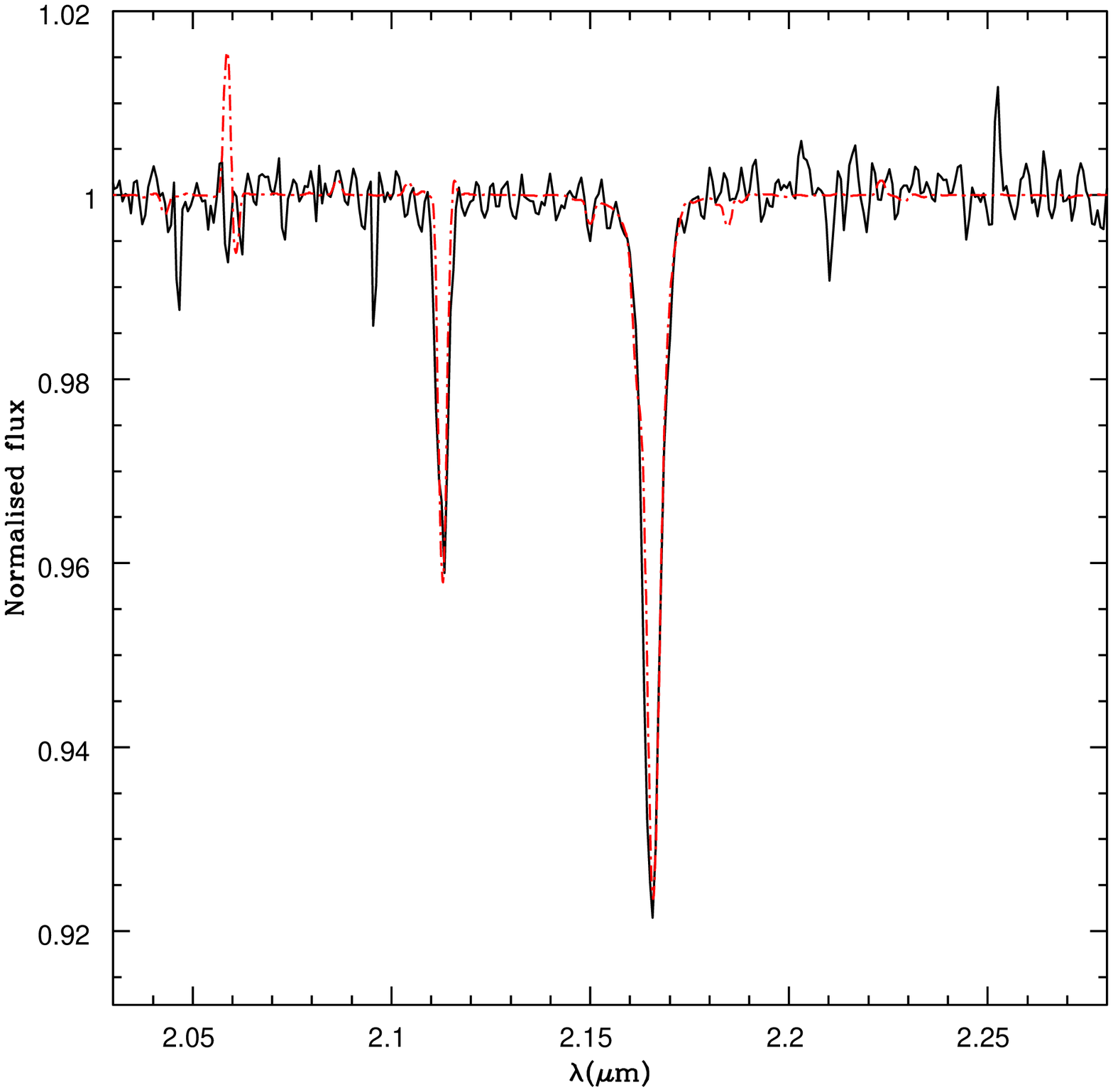}
\includegraphics[width=6cm]{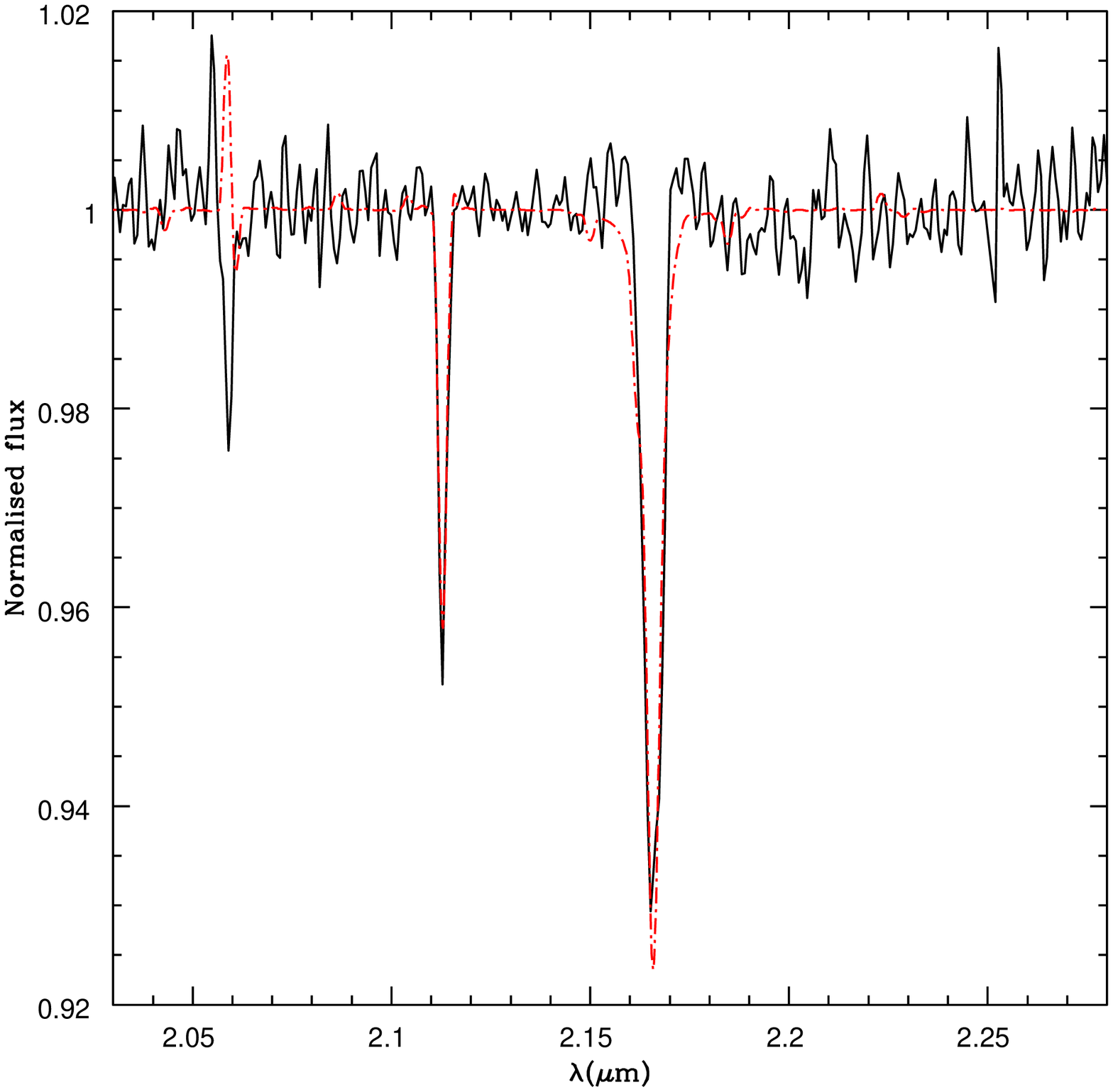}
\includegraphics[width=6cm]{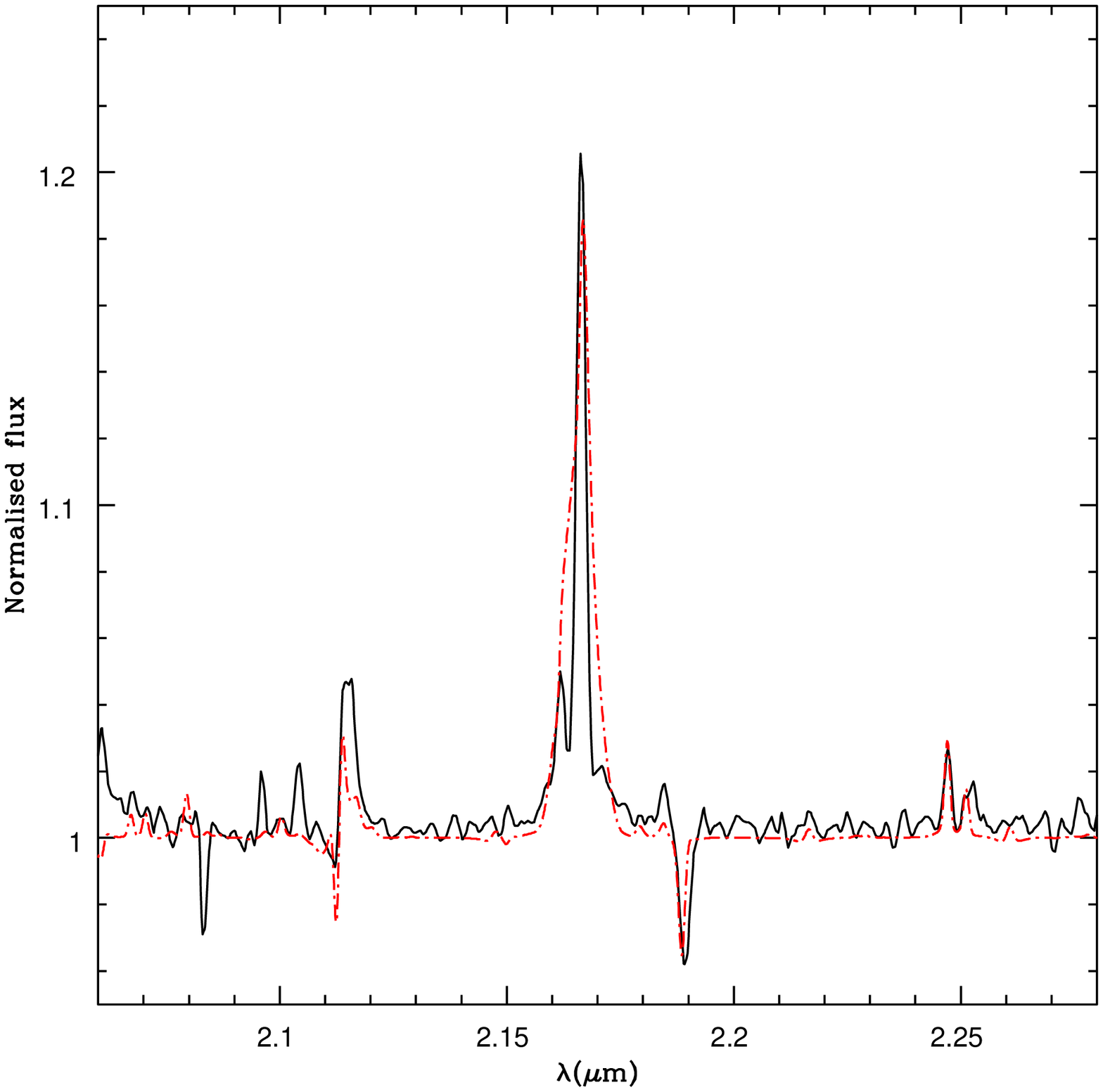}
\caption{Comparison of our best fit model (red dotted lines) to the observations (black solid lines) of two O8-9V (VVV\,CL009-05, VVV\,CL009-08, left an middle panels respectively) and one OIf/WN7 star (VVV\,CL009-6, right panel) in VVV\,CL009.}
\label{fit_vc09}
\end{center}
\end{figure}

Concerning VVV\,CL009-6, a fit of moderate quality is obtained (see Fig.\,\ref{fit_vc09}, right panel). Fitting the narrow and relatively strong emission lines requires a moderate terminal velocity and a quite high mass loss rate (see Table~\ref{tab_param}). We suspect that part of the emission observed in Br$\gamma$ is actually due to residual nebular emission. The mass loss rate we determine, mainly based on Br$\gamma$, should thus be regarded as an upper limit. The \ion{N}{iii} doublet at 2.24\mum\ is clearly detected and allows to derive a high nitrogen abundance. An upper limit on the carbon content is obtained from the absence of \ion{C}{iv} emission at 2.04-2.08\mum. 

The observed spectra of the two O dwarfs are very similar. Our models reproduce the observed spectra quite well (Fig.\,\ref{fit_vc09} left and central panel). In both cases, the intensity and profile of Br$\gamma$ are well accounted for, as well as the 2.11$\mu$m line complex. We found very similar effective temperatures and mass loss rates for both stars. As no carbon nor nitrogen line is visible in the spectra we have assumed solar abundances.

\begin{sidewaystable}
\begin{center}
\begin{threeparttable}
\caption{Stellar and wind parameters} \label{tab_param}
\begin{tabular}{llllllllllllll}
\hline
Star    &    ST		& T$_*$	& \teff	& \logg	&  log(L/L$_{\odot}$)    & log($\dot{M}$) & \vinf & H	& He	&   C           &    N	       \\ 
        &      		& [kK]	& [kK]	&       &     		&            	  &[\kms]	&	&       & [$10^{-4}$]    &  [$10^{-3}$]     \\ 
\hline
VVV\,CL009-5	& O8-9V   	& 32.1 	& 32.0	& 4.00$^*$  & 4.68 $\pm$ 0.12	& -7.0*  & 2735*    & 0.70*	& 0.28*	&   -   &  -     \\
VVV\,CL009-6	& OIf/WN7  	& 36.0 	& 34.0	& 3.25$^*$  & 5.25 $\pm$ 0.11	& -5.8   & 1200*    & 0.70	& 0.28	&  0.9  &  4.0   \\
VVV\,CL009-8	& O8-9V         & 32.1	& 32.0	& 4.00$^*$  & 4.55 $\pm$ 0.12	& -7.0*  & 2735*    & 0.70*	& 0.28*	&   -   &  -      \\
%\hline  
%VVV\,CL036-4	& B2-3I   	&$<$29.1 & $<$26.0& 2.75$^*$ & 5.27 $\pm$ 0.15	& -6.5*  & 2000*    & 0.70*	& 0.28*	& -     &  -     \\
%VVV\,CL036-9	& WN6     	& 59.0	& 50.0	& 3.50$^*$   & 5.-- $\pm$ 0.08	& -5.0	 & 2750	    & 0.01      & 0.76	& 1	&  32 	 \\
\hline
VVV\,CL073-2	& O4-6If+/WN9   & 35.4  & 34.0	& 3.30$^*$   & 5.30 $\pm$ 0.11	& -5.7   & 2000	    & 0.70	& 0.28  & 4.0   &  5.0   \\
VVV\,CL073-4	& WN7	  	& 42.2 	& 34.0	& 2.90$^*$   & 5.70 $\pm$ 0.11	& -4.8   & 1400	    & 0.26	& 0.73	& $<$3  &  5     \\
\hline                                                                                                                                         
VVV\,CL074-1	& O8.5I		& 31.3	& 31.0	& 3.50$^*$   & 5.21 $\pm$ 0.12	& -6.0   & 2000	    & 0.70	& 0.28	& $<$30 &  8.0      \\
%VVV\,CL074-2	& WC8           & 4	& 3  & 3.00$^*$   & 5.-- $\pm$ 0.12	& -5.5   & 3000	    & 0.00	& 0.  & 	&  	  \\
VVV\,CL074-3    & WN8           & 32.4  & 31.0  & 3.50$^*$   & 5.70 $\pm$ 0.13  & -4.7   & 600      & 0.13      & 0.85  & 1.3  &  11.9  \\
VVV\,CL074-5	& O8.5I		& 33.3	& 33.0	& 3.50$^*$   & 5.44 $\pm$ 0.12	& -5.8   & 2000	    & 0.70	& 0.28	& $<$4  &  4  	 \\
VVV\,CL074-7	& O8.5I	 	& 33.4	& 33.0	& 3.50$^*$   & 5.11 $\pm$ 0.12	& -5.9   & 2000	    & 0.70	& 0.28	& $<$4  &  2     \\
VVV\,CL074-9	& O4-6If+	& 34.4	& 34.0	& 3.50	     & 5.52 $\pm$ 0.11	& -5.7   & 2000	    & 0.70	& 0.28	& $<$8  &  7   	 \\
\hline
VVV\,CL099-5	&  WN6		& 62.1	& 51.2	& 4.00$^*$   & 5.10 $\pm$ 0.10	& -4.6   & 2750	    & 0.00	& 0.96	& 0.6 	&  10 	 \\
VVV\,CL099-7	&  WC8     	& 47.4	& 45.1	& 3.50$^*$   & 5.50 $\pm$ 0.10	& -4.7   & 2200	    & 0.00	& 0.86	& 1300  &  0.2  \\
\hline
\end{tabular}
\begin{tablenotes}
  \footnotesize
  \item * $=$ adopted.
  \item Abundances are given in mass fraction.
\end{tablenotes}
\end{threeparttable}
\end{center}
\end{sidewaystable}

%%###############################################
\subsection{VVV\,CL073 cluster}

\begin{figure}
\begin{center}
\includegraphics[width=6cm]{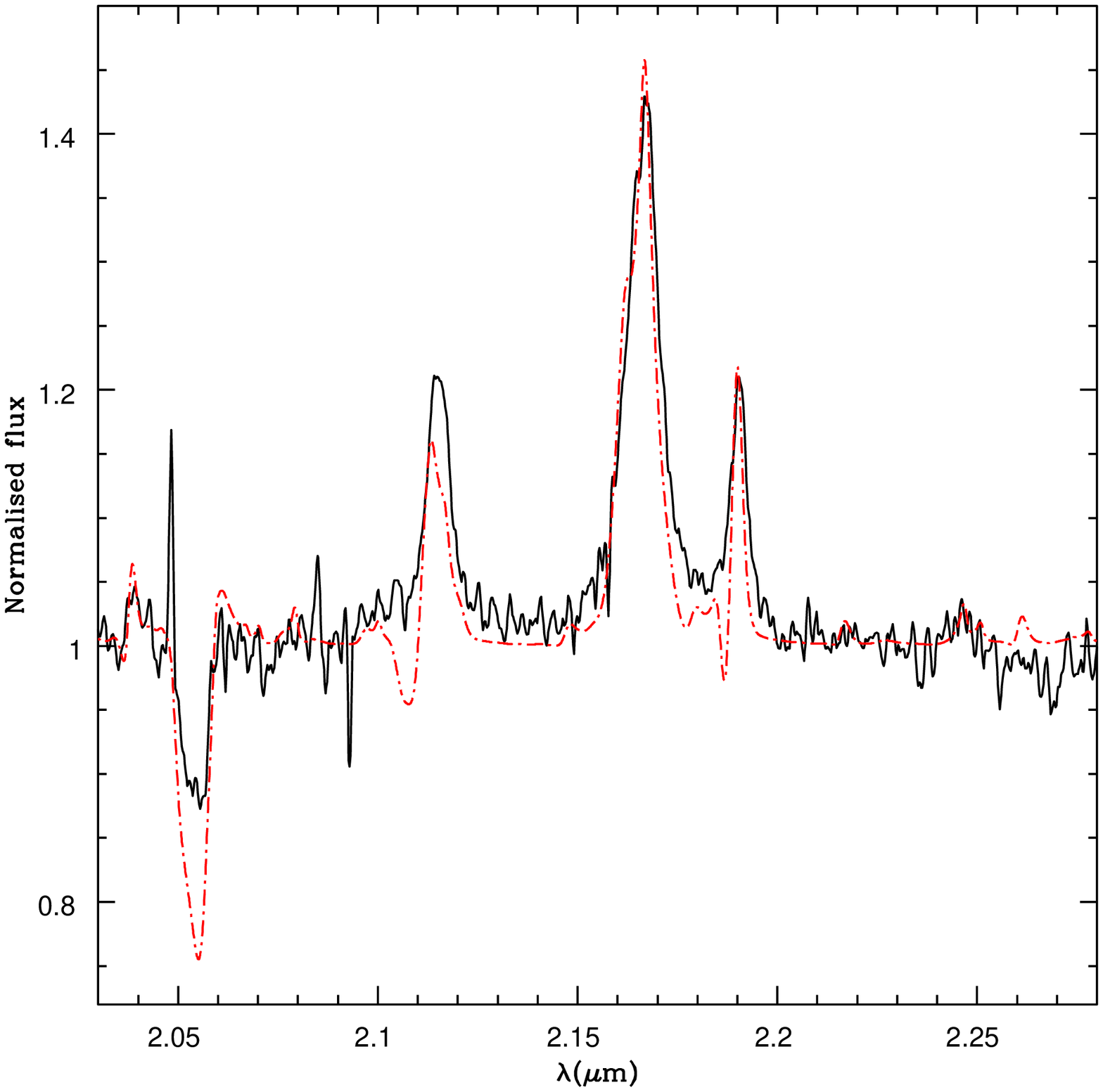}
\includegraphics[width=6cm]{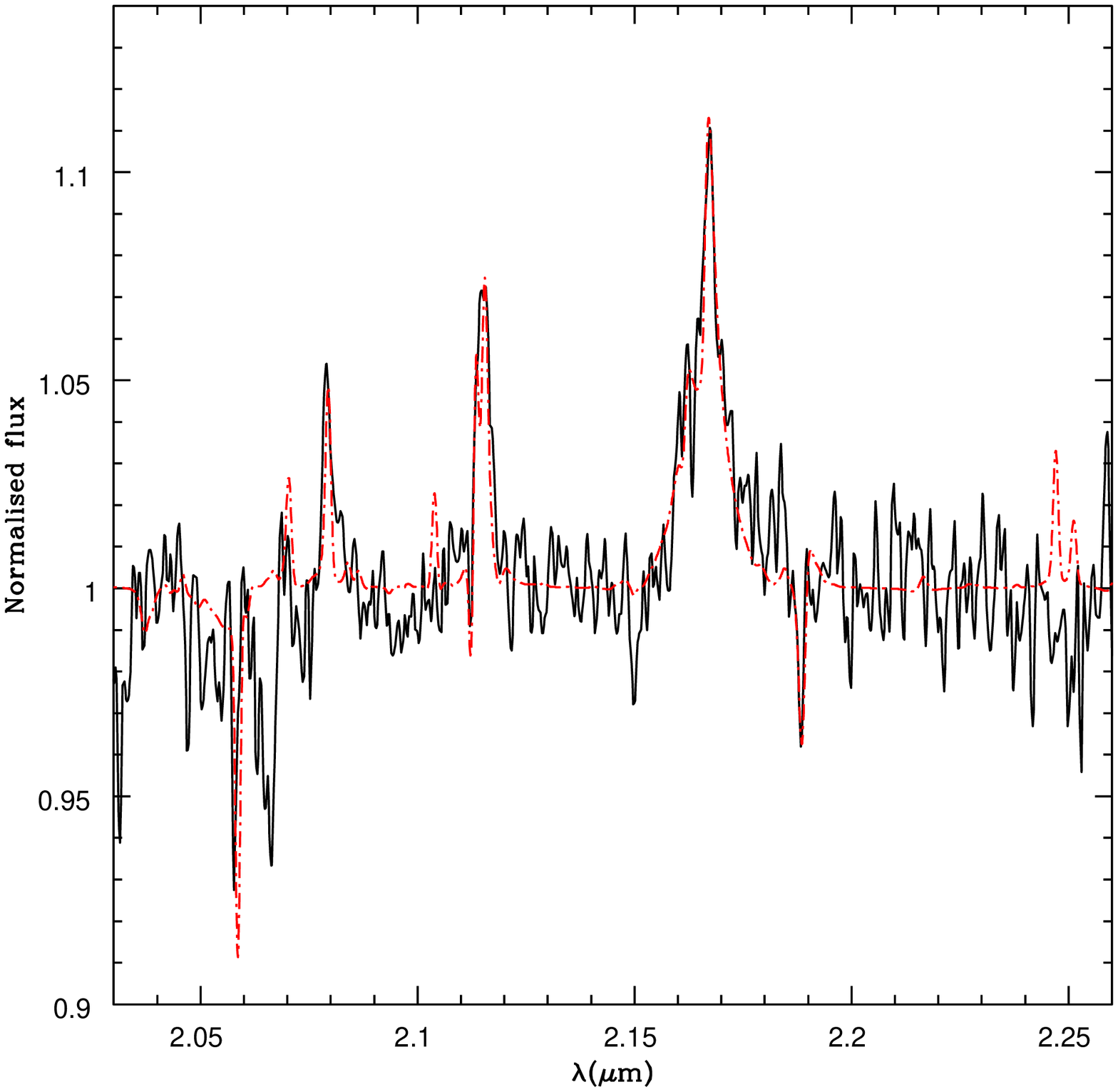}
\caption{Comparison of our best fit model (dotted red line) to the observed spectra (solid black line) of the WN7 star VVV\,CL073-4 (left panel) and of the O4-6If+/WN9 star VVV\,CL073~2 (right panel).}
\label{fit_vc73}
\end{center}
\end{figure}

We have analyzed the WN7 star VVV\,CL073-4 and O4-6f+/WN9 object VVV\,CL073-2. A good fit of the observed spectra was obtained (see Fig.\ \ref{fit_vc73}). For both stars, the ratio of the \ion{He}{i} to \ion{He}{ii} lines is fairly well accounted for, ensuring a good temperature determination as well as reasonable constraints on the helium content. VVV\,CL073-4 is significantly helium enriched, as expected from its spectral class. On the other hand, fitting the helium and hydrogen spectrum of VVV\,CL073-2 does not require He/H different from the solar value. Both stars are depleted in carbon (only an upper limit on C/H is derived for VVV\,CL073-4) and enriched in nitrogen. The mass loss rate of the WN7 star is much higher than that of the O4-6f+/WN9. Overall, VVV\,CL073-4 appears to be in a more advanced evolutionary state than VVV\,CL073-2.

%%###############################################
\subsection{VVV\,CL074 cluster}

This cluster contains the largest number of massive stars with good spectroscopic data. We determined the stellar parameters of one WN8, one O4-6If+ and three O8.5I stars.

\begin{figure}
\begin{center}
\includegraphics[width=0.32\textwidth]{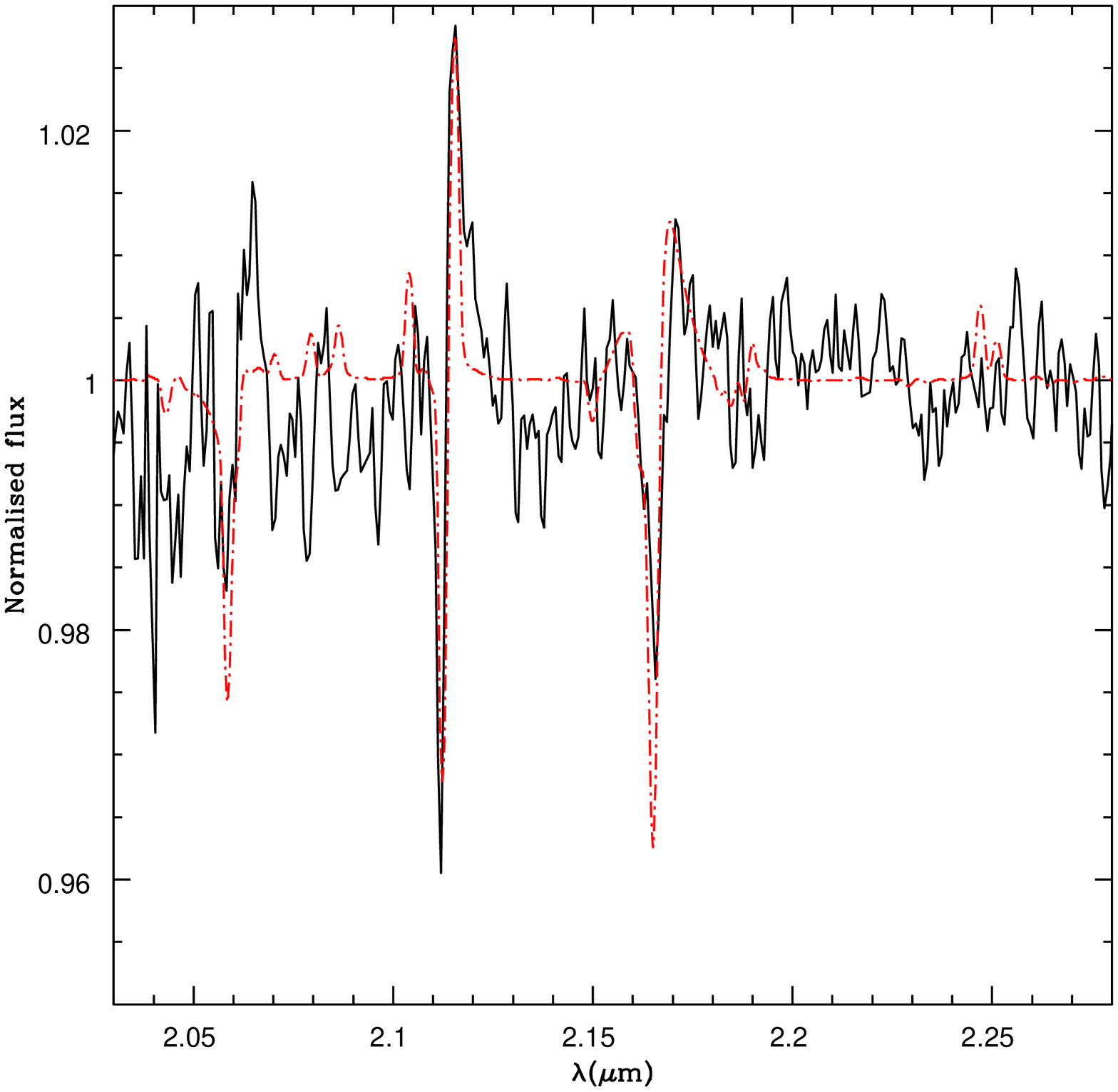}
\includegraphics[width=0.32\textwidth]{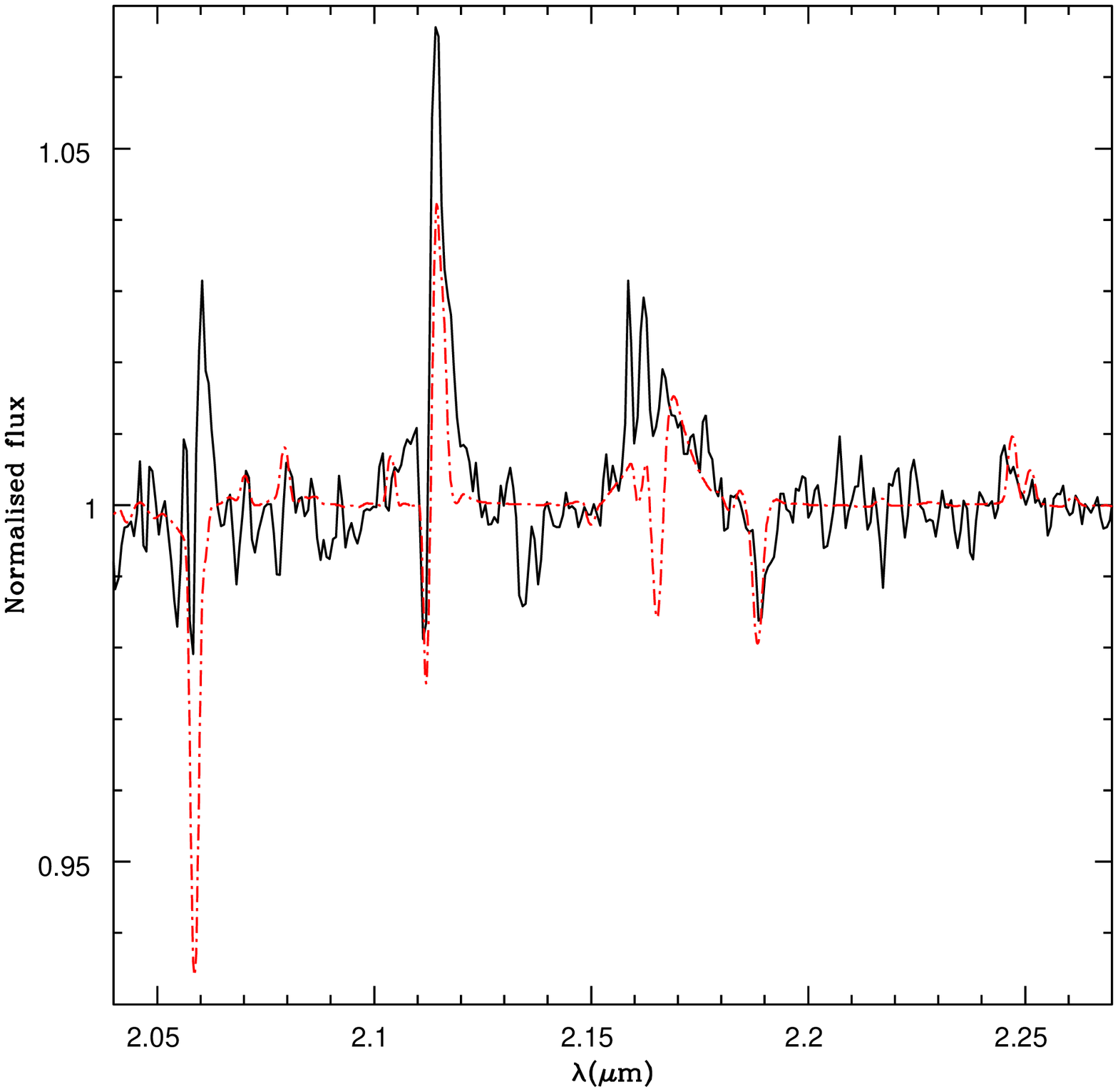}
\includegraphics[width=0.32\textwidth]{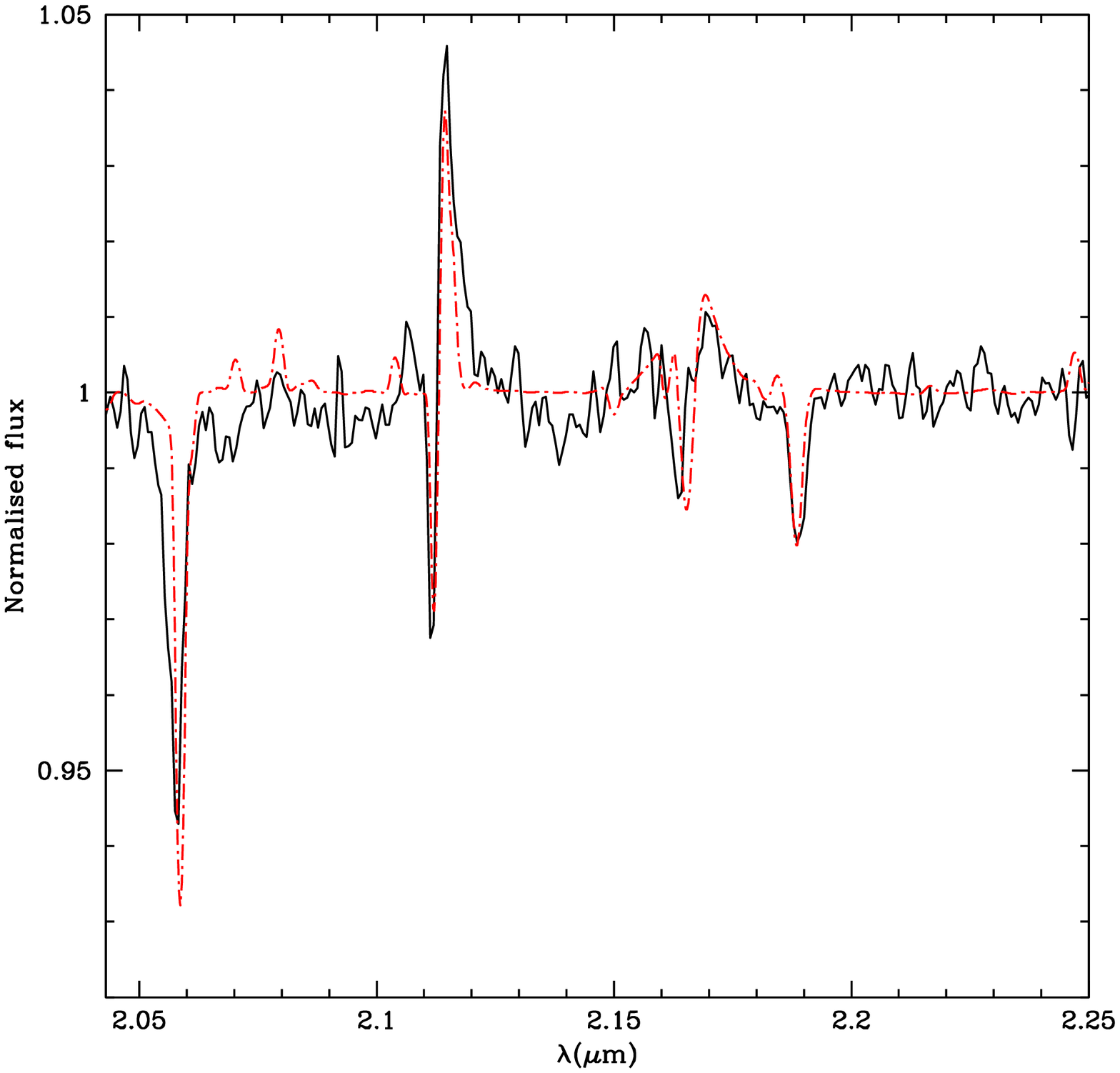}\\
\includegraphics[width=0.32\textwidth]{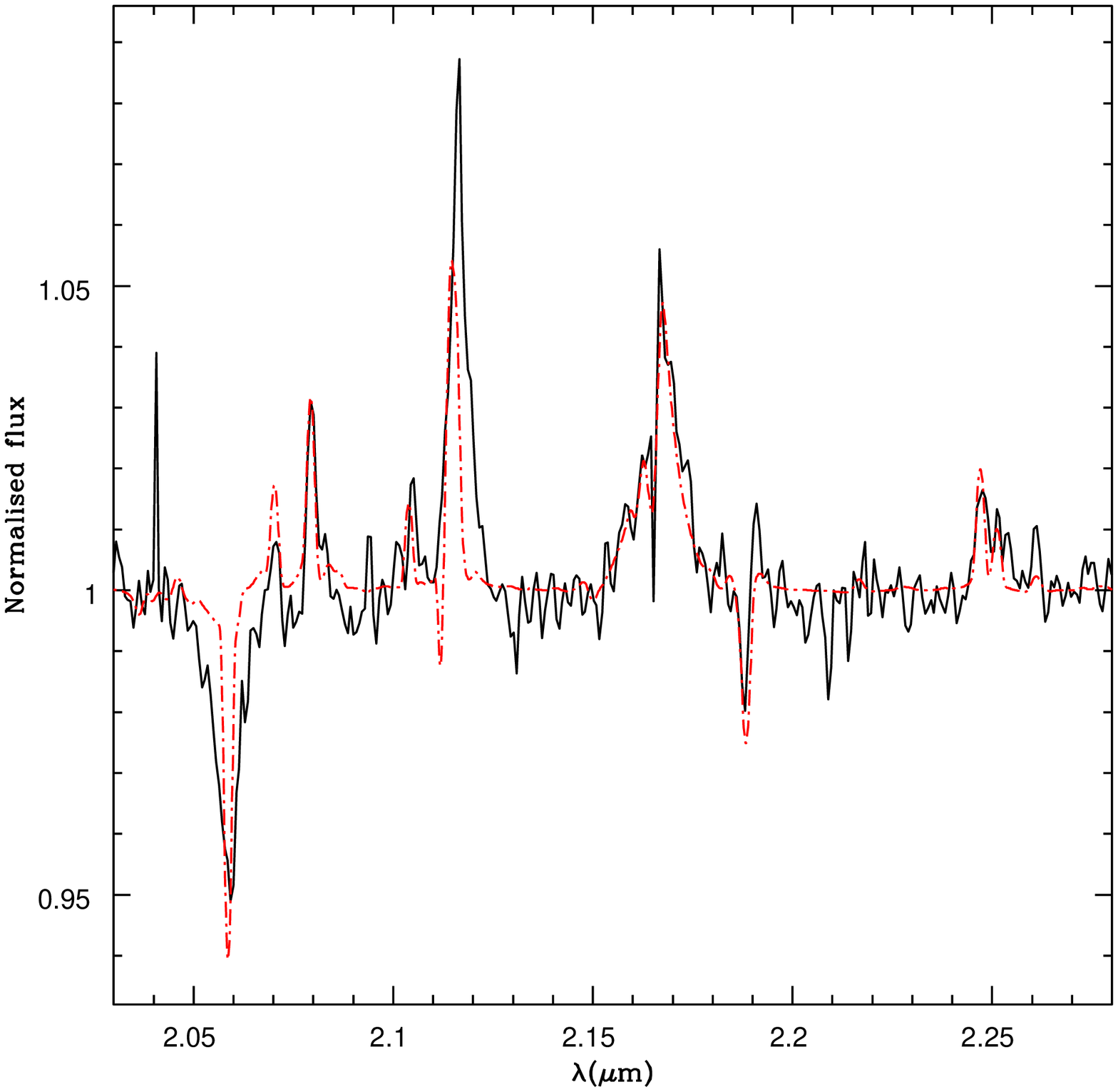}
\includegraphics[width=0.32\textwidth]{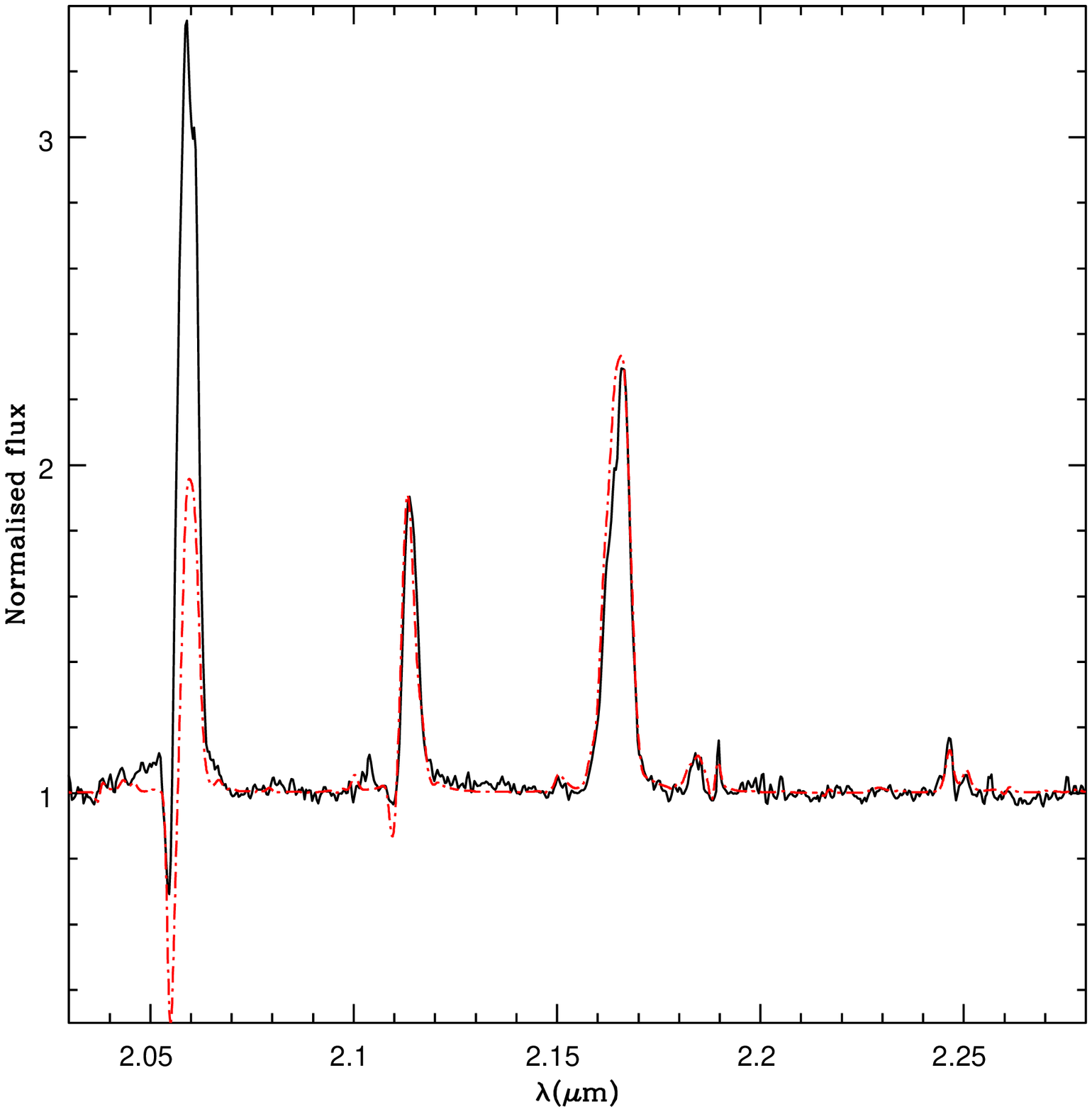}
\caption{Comparison of our best fit model (in dotted red lines) to the observation  (in solid black lines) of three O8.5I stars (VVV\,CL074-1, VVV\,CL074-5, VVV\,CL074-7, top panel), a WN7-O4-6If (VVV\,CL074-9) and a WN8 (VVV\,CL074-3) -- bottom panel.}
\label{fit_vc74}
\end{center}
\end{figure}

The spectra of the three O8.5I are very similar. For each of them, our fits reproduce the different H and He lines present in the K band (Fig.\,\ref{fit_vc74}\, top panel). The small differences in the \ion{He}{i}/\ion{He}{ii} ratio translate into differences of few thousands of Kelvin in their effective temperature. The different intensity of Br$\gamma$ also revealed slight differences in mass loss rates. Upper limits in the carbon content could be obtained from the strength of the \ion{C}{iv} lines. The nitrogen doublet at 2.24$\mu$m provided indications of nitrogen enrichment. 

Concerning the O4-6If+ star, the quality of the fit is good. Most of the lines are reasonably well reproduced. A rather strong mass loss rate explains the emission of Br$\gamma$. We found a carbon depletion and a strong nitrogen enrichment. 

The spectrum of the WN8 star is well fitted by our best model. The only exception is the \ion{He}{i} 2.058$\mu$m line which is known to depend on details of the modelling and atomic physics \citep{najarro97,najarro06,martins07}. The strong \ion{He}{i} emission lines dominate the spectrum due to a large helium content and a strong mass loss rate. The presence of \ion{N}{iii} lines indicates nitrogen enrichment. The absence of carbon lines is a sign of carbon depletion. The width of the emission lines indicates a low terminal velocity.

%%###############################################
\subsection{VVV\,CL099 cluster}

\begin{figure}
\begin{center}
\includegraphics[width=6cm]{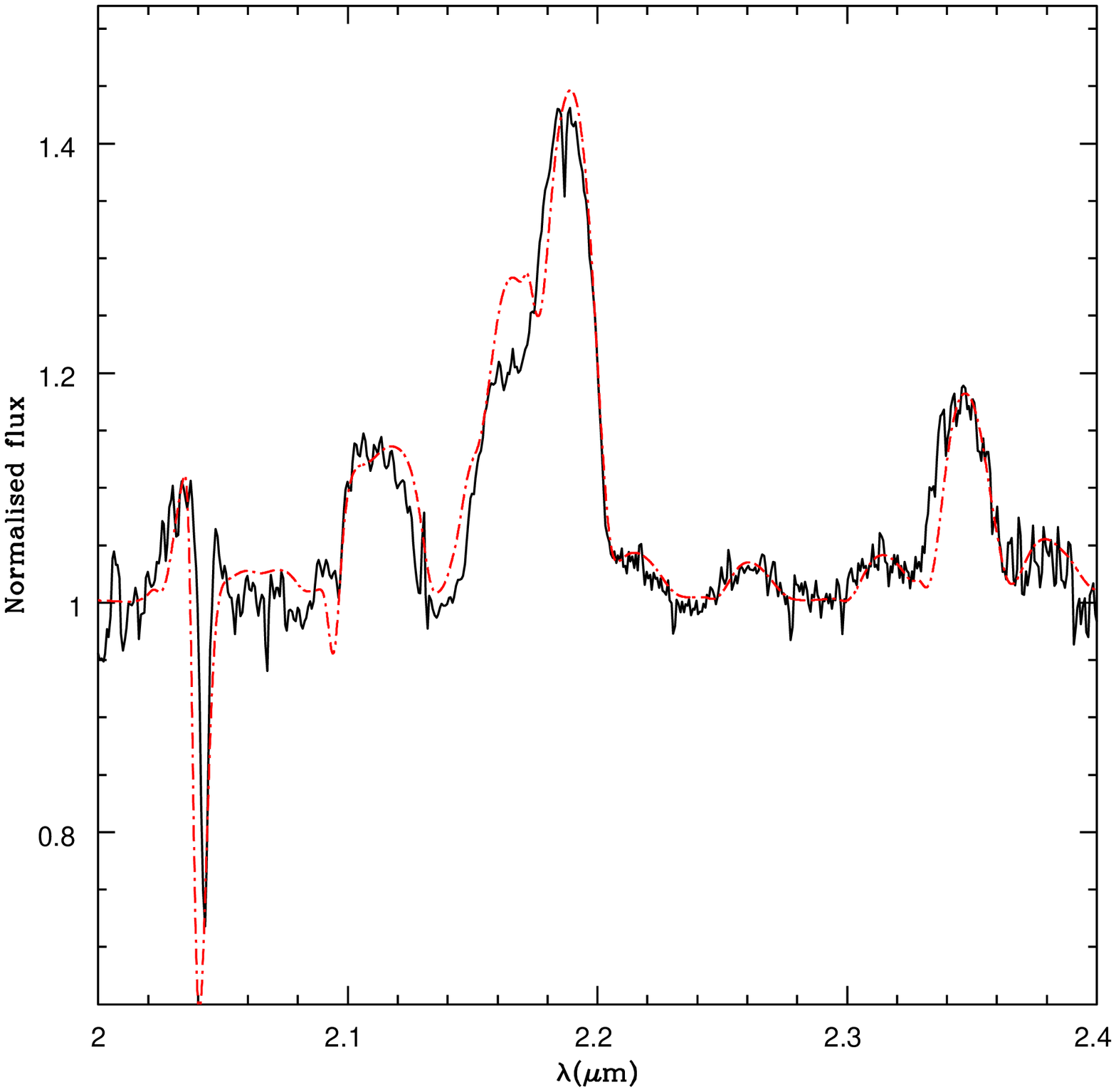}
\includegraphics[width=6cm]{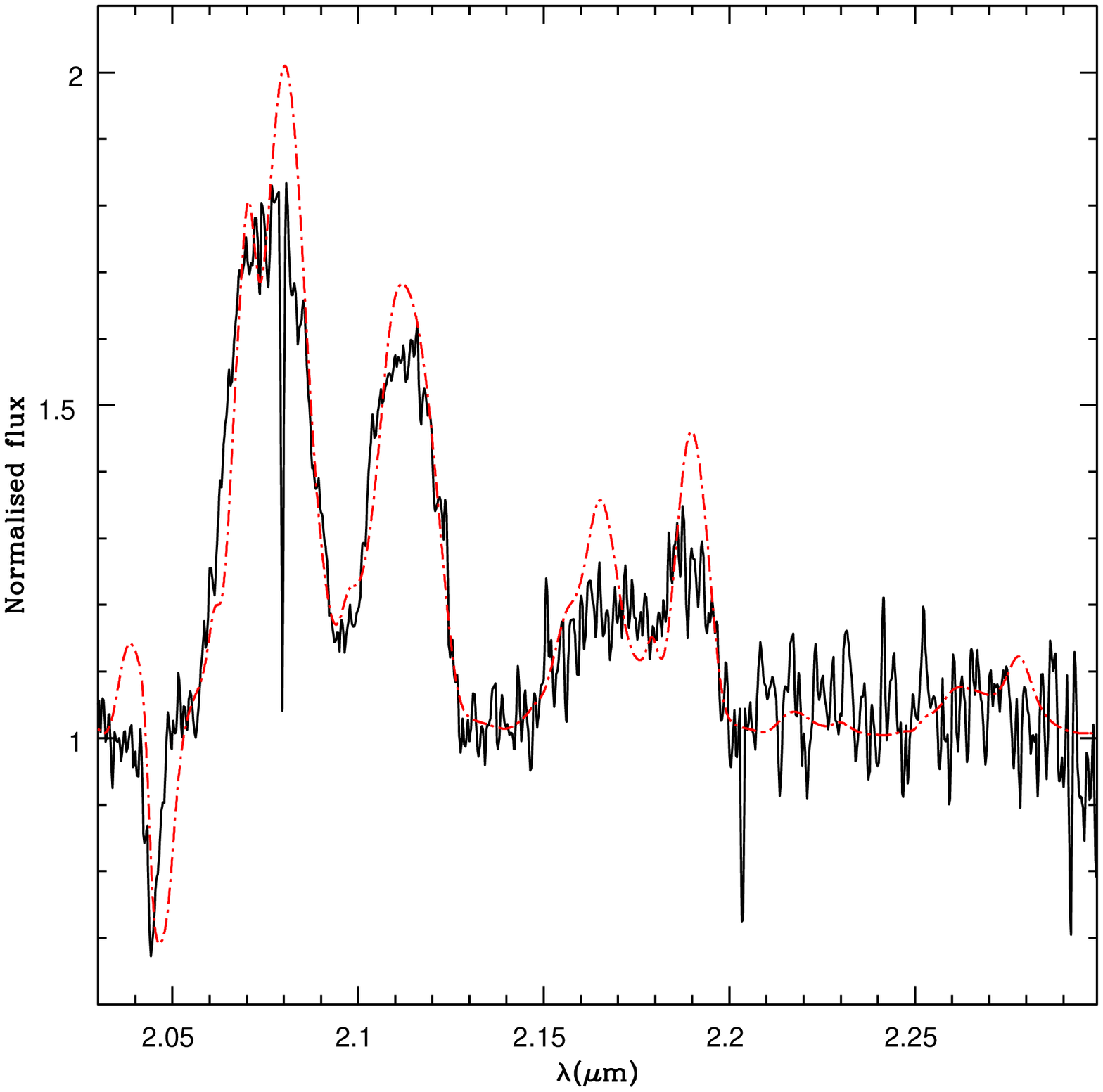}
\caption{Comparison of our best fit model (dotted red line) to the observed spectrum (solid black line) of VVV\,CL099-5 (WN6, Left panel) and VVV\,CL099-7 (WC8, right panel).}
\label{fit_vc99}
\end{center}
\end{figure}

A WN6 star (VVV\,CL099-5) and a WC8 star (VVV\,CL099-7) have been detected in this cluster. Our best fit models are presented in Fig.\,\ref{fit_vc99}.

Concerning the WN6 star, our best fit model provides a very good match to the observed spectrum. We observe very intense and broad emission lines of \ion{He}{i}, \ion{He}{ii} and broad \ion{N}{iii} lines. These different lines have been used to determine the effective temperature, a strong mass loss rate and high terminal velocity, and a strong nitrogen enrichment. 
 
In the case of the WC8 star, our fit is reasonable, although the intensity of the helium lines around 2.2 $\mu$m is still a little high. Reducing their intensity requires to reduce the luminosity and/or the mass loss rate, which translates in a too large absolute magnitude. Reducing the helium content does not help either.
From the \ion{c}{iii} and \ion{c}{iv} lines, we found a significant carbon enrichment consistent with the spectral type of the star.

%%###############################################
%%###############################################
\section{Discussion}
\label{s_disc}

\begin{figure}
\begin{center}
\includegraphics[width=0.49\textwidth]{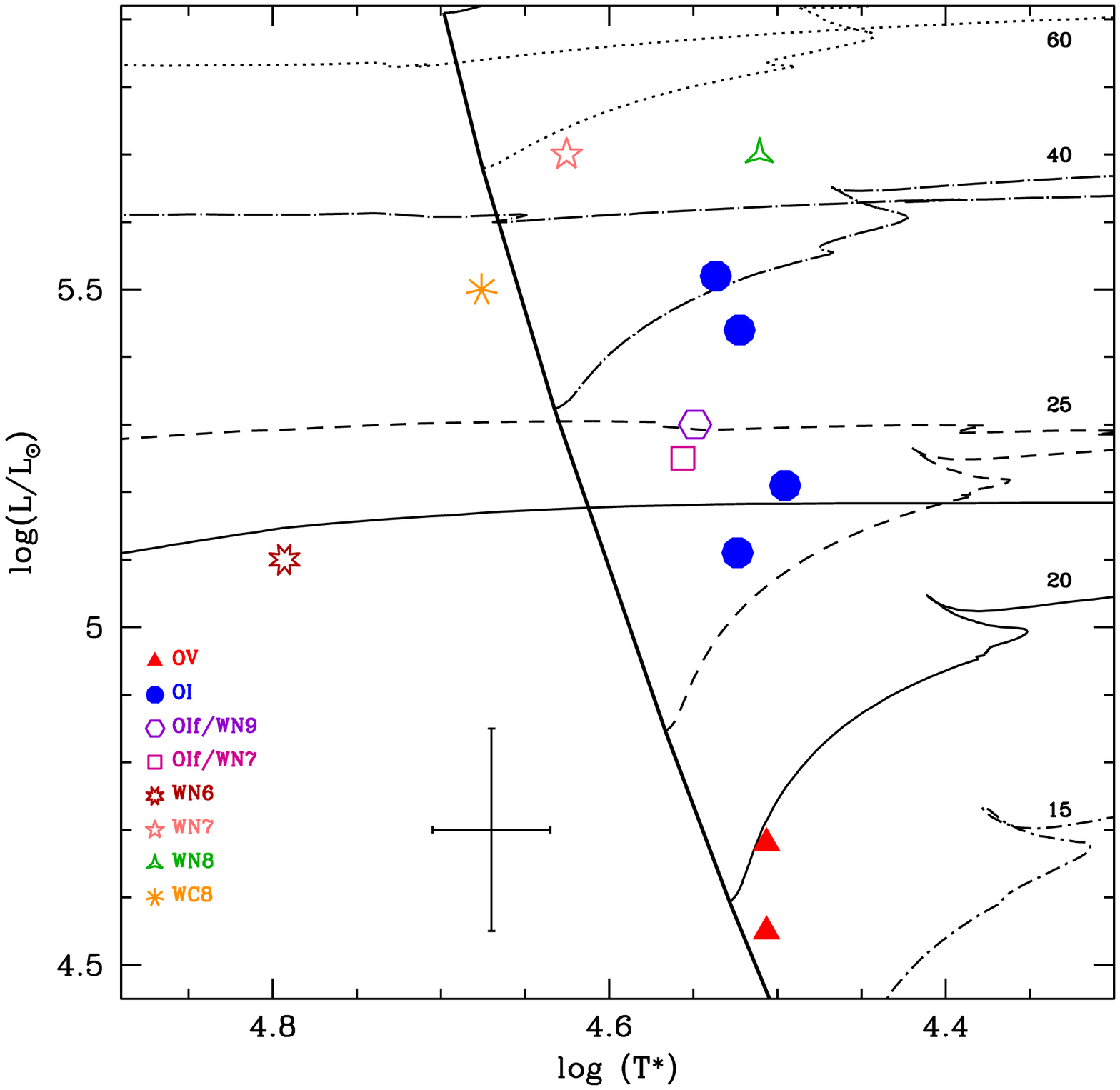}
\includegraphics[width=0.49\textwidth]{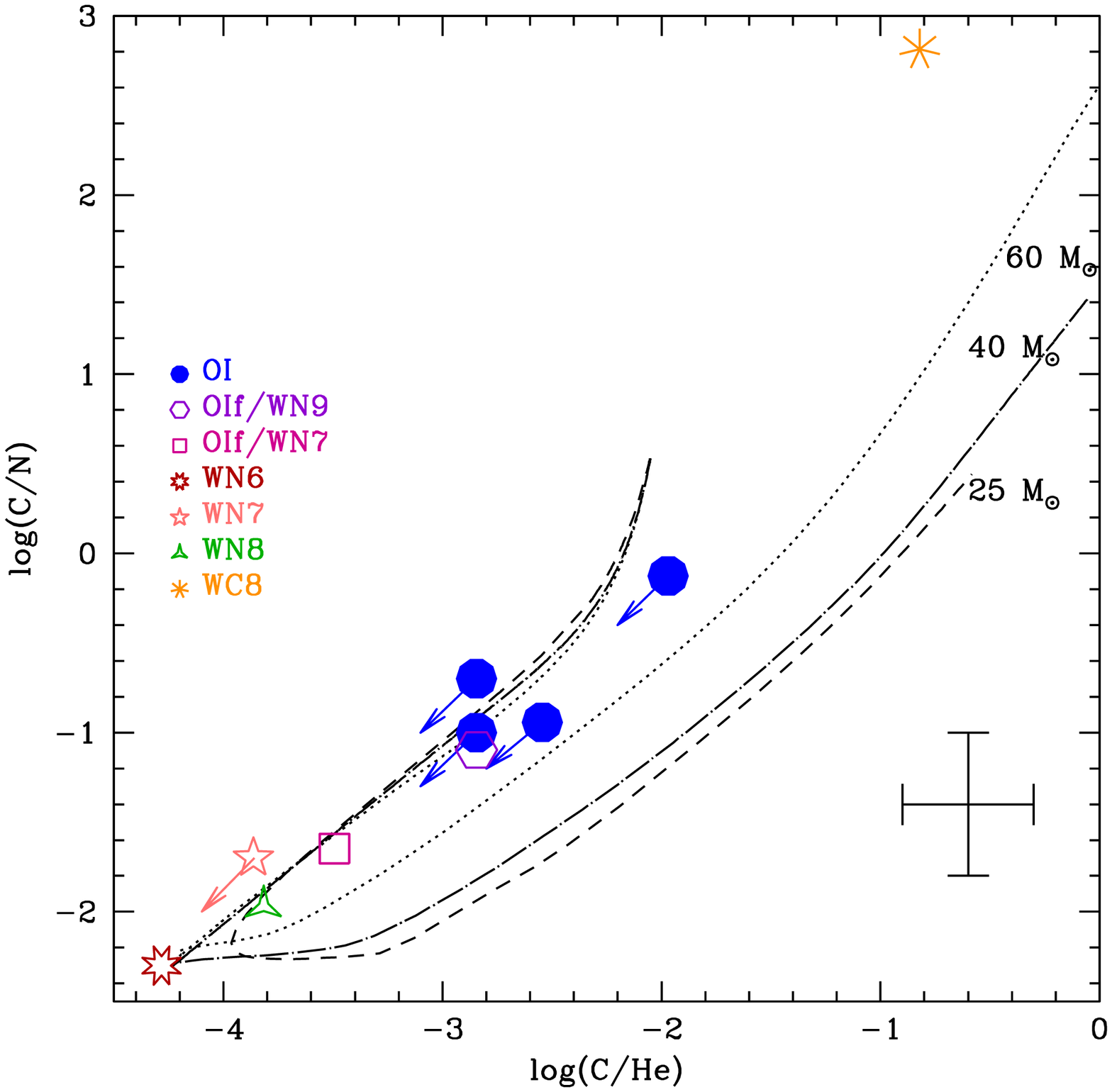}\\
\includegraphics[width=0.49\textwidth]{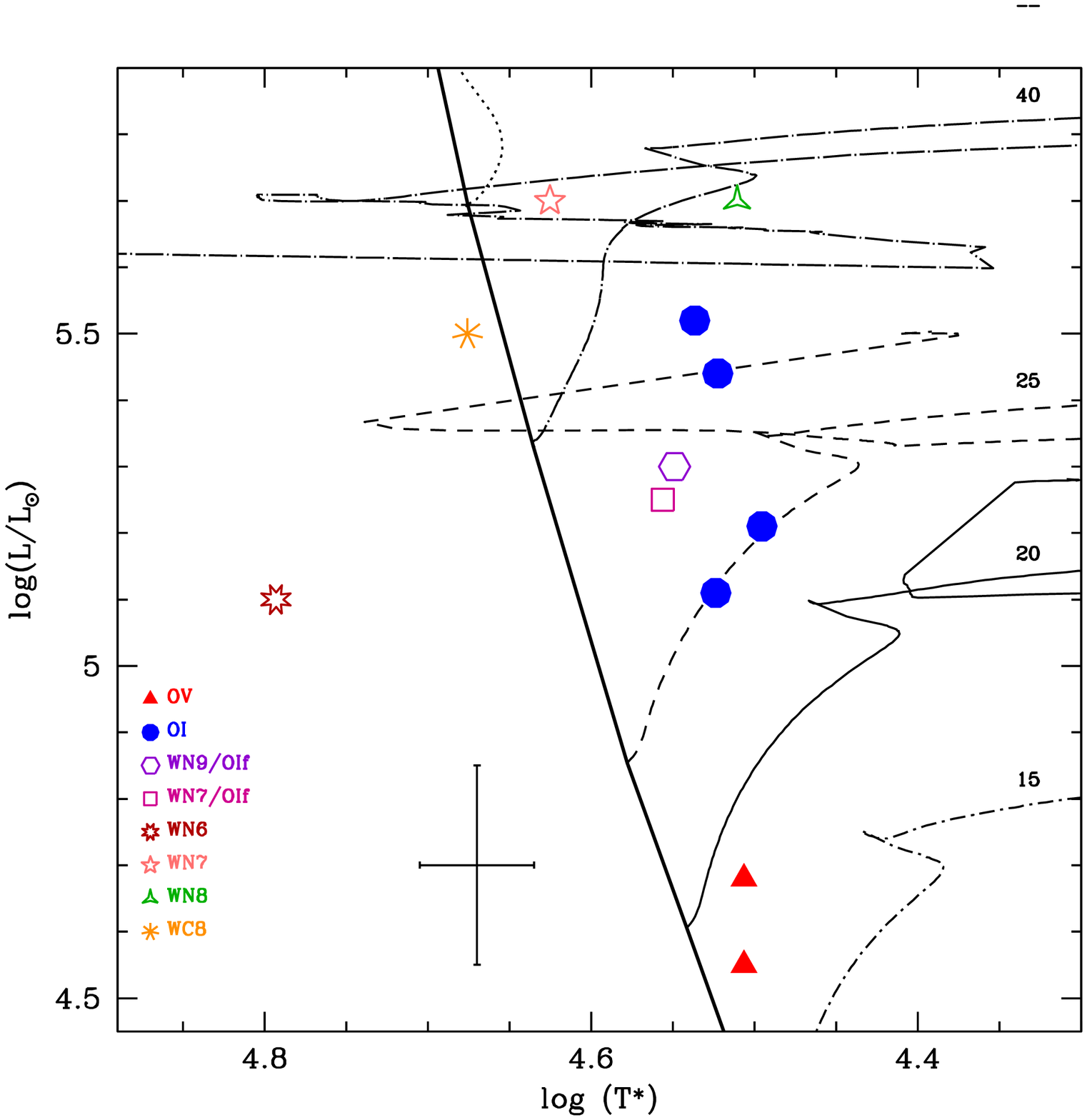}
\includegraphics[width=0.49\textwidth]{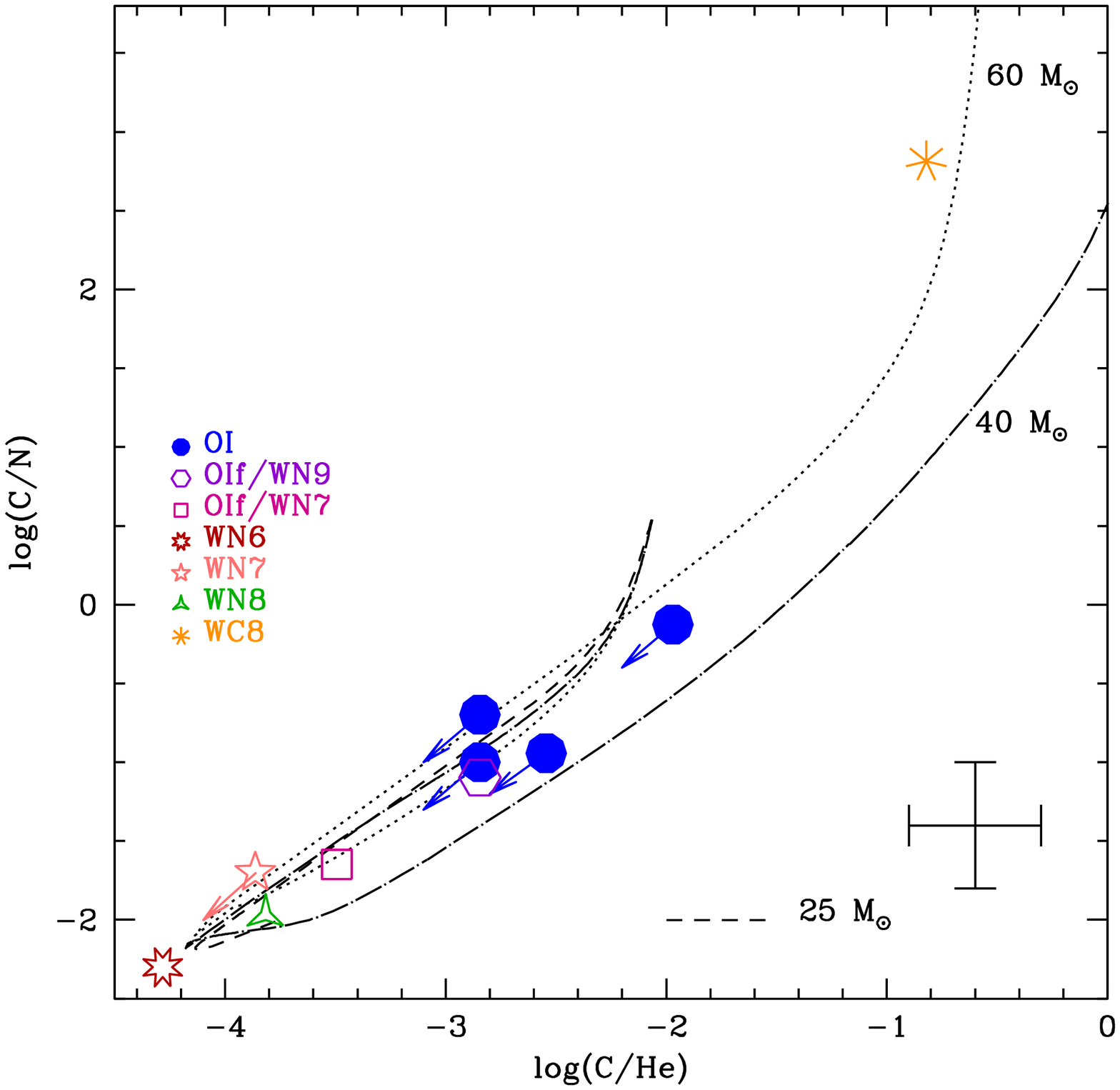}
\caption{Hertzsprung-Russell diagram (left panels) and log (C/N) -- log (C/He) diagram (right panels) with evolutionary tracks are from \citet{cl13} (top panels) and \citet{ek12} (bottom panels). Tracks include rotation. Different symbols correspond to different types of stars.}
\label{hrclusters}
\end{center}
\end{figure}

Given the small number of sources per cluster, we have grouped all stars in the same HR diagram and log(C/N) -- log(C/He) diagram (Fig.\ \ref{hrclusters}). We thus discuss our results regardless of the cluster. We show these diagrams with the evolutionary tracks of \citet{cl13} and \citet{ek12}. 

In the HR diagram, the two O dwarfs are located at relatively low luminosity, close to the zero-age main sequence (ZAMS). The O supergiants are more luminous and also farther away from the ZAMS. They are actually close to the terminal-age main sequence. Our results thus confirm the trend expected from spectral types (supergiants are more evolved in the HR diagram than dwarfs). The O dwarfs have initial masses in the range 15-20 \msun\ while the O supergiants are more massive (25-40 \msun). In the abundance diagram (right panels of Fig.\ \ref{hrclusters}), the O supergiants are located close to (but not exactly at) the starting point of the evolutionary tracks, indicating that they are chemically little evolved (compared to other stars, see below). 

The two objects with a transition classification OIf/WN (VVV\,CL009-6 and VVV\,CL073-2) are located at the same place as the other O supergiants in the HR diagram. Their transition classification is justified by the strength of their emission lines. Their mass loss rates are also similar to those of O supergiants. 
In the log(C/N) versus log(C/He) diagram, one transition object is close to the O supergiants while the other is slightly more evolved. One may suspect that these transition objects are 25 \msun\ stars having evolved to the red part of the HRD and now moving back to the blue. But at their position in the HR diagram, the 25 \msun\ track of \citet{cl13} predicts a significant helium enrichment (Y=0.43) which is not consistent with our determinations. We thus favor the interpretation that VVV\,CL009-6 and VVV\,CL073-2 are normal O supergiants perhaps, in one case, slightly more evolved than O supergiants.

The four Wolf-Rayet stars of our sample occupy different regions of the HR diagram and of the log(C/N) - log(C/He) diagram. They are more luminous or hotter than the O stars. A higher luminosity is indicative of a higher initial mass (40-60 \msun\ for the WN7 and WN8 stars according to \citealt{cl13}; $\sim$40 \msun\ according to \citealt{ek12}). The WN6 star (VVV\,CL099-5) can be explained by the 20 and 25 \msun\ tracks of \citet{cl13}. However, no model from the grid of \citet{ek12} is able to account for the temperature and luminosity of VVV\,CL099-5. The WC8 star is probably the descendent of an O star with an initial mass between 25 and 40 \msun\ (depending on the set of tracks). 

In the log(C/N) -- log(C/He) diagram, all WN stars are located in the lower left corner, where nitrogen/helium enrichment and carbon depletion are the strongest. 
The position of the WN stars in this diagram is well reproduced by the evolutionary models of \citet{cl13} and \citet{ek12}. We note that evolutionary tracks with different initial masses follow the same path in the log(C/N) -- log(C/He) diagram, at least in the lower left part. This indicates that CNO equilibrium has been reached and that we see its products at the surface of the stars. This is independent of the mixing processes (and their dependence on initial mass) and of the numerics of the code used to produce the tracks (the tracks from \citealt{cl13} and \citealt{ek12} predict consistent evolution). 

The WC8 star is located at the extreme opposite of the log(C/N) - log(C/He) diagram, where carbon enrichment is very strong. Our results thus indicate that WC stars are in a more advanced state of evolution than WN stars of type WN6 to WN8 \citep[see also][]{martins07}. The Wolf-Rayet stars of our sample are also more chemically evolved than O-type stars. These results confirm the predictions of evolutionary models: WN stars show products of strong CNO processing; they are more chemically evolved than O supergiants. The WC star shows the products of helium burning, with a high carbon content. It is thus in a more advanced state of evolution than the WN stars.

Inspection of Table \ref{tab_param} reveals that the mass loss rates of the Wolf-Rayet stars are on average 10 times higher than those of O stars. Indeed, we find log($\dot{M}$) between -4.8 and -4.6 for the former, while for the latter the values are between -5.7 and -6.0. Even with relatively large uncertainties in our determinations, the difference is significant. It is also consistent with what is usually found for massive stars. Our mass loss rates are in very good agreement with the average values reported by \citet{crowther07} : \mdot\ = $10^{-4.8..-4.7}$ \myr\ for WN6-8 stars. Similarly, the mass loss rates reported by \citet{hamann06} for late-type WN stars in the luminosity range 5.6 $<$ \lL\ $<$ 5.8 are consistent with our measurements (see their Fig.~6). Mass loss rates for Galactic O stars with luminosities around $10^{5.5}$ L$_{\odot}$ are usually close to $10^{-5.6..-6.2}$ \myr\ \citep{mokiem05}, similar to what we find.

%%###############################################
%%###############################################
\section{Conclusion}
\label{s_conc}

We have presented the analysis of massive stars in four young massive clusters recently discovered by the VVV survey. Based on near-infrared spectroscopy, we have determined the fundamental parameters of twelve stars with various spectral types: O dwarfs, O supergiants, WN and WC stars. Atmosphere models computed with the code CMFGEN have been used. We have also determined surface abundances (H, He, C and O) for a sub-set of these stars.
We have shown that the Wolf-Rayet stars are more chemically evolved than the O supergiants, with a clear sequence of chemical enrichment when moving from O supergiants to WN and WC8 stars. Mass loss rates among Wolf-Rayet stars are a factor of 10 larger than for O stars, in agreement with previous findings.
 
The observation and subsequent analysis of additional targets in these clusters will help identify the main sequence, and thus the clusters' turn-off and detailed evolutionary sequence. This is crucial to place constraints on the clusters' ages and thus to test quantitatively the predictions of evolutionary models.

%%###################################################
\vspace{0.3cm}

\noindent \textbf{Acknowledgements}\\
FM and AH thank the Agence Nationale de la Recherche for financial support (grant ANR-11-JS56-0007). AH is supported by the grant 14-02385S from GA \v{C}R. JB are supported by Ministry of Economy, Development, and Tourism's Millennium Science Initiative through grant IC120009, awarded to The Millennium Institute of Astrophysics, MAS.  J.B, RK and SR acknowledge funds from Fondecyt Regular N$^{o}$1120601, 1130140 and 3140605.



\vspace{1.0cm}

\noindent \textbf{References}\\

\end{document}